\documentstyle[12pt]{article}

\catcode`\@=11
\def\maketitle{\par
\begingroup
\def\thefootnote{\fnsymbol{footnote}}
\def\@makefnmark{\hbox
 to 0pt{$^{\@thefnmark}$\hss}}
\if@twocolumn
\twocolumn[\@maketitle]
\else \newpage
\@maketitle \fi
\thispagestyle{empty}
\@thanks
\endgroup
\setcounter{footnote}{0}
\let\maketitle\relax
\let\@maketitle\relax
\gdef\@thanks{}
\gdef\@author{}
\gdef\@title{}
\gdef\@preprint{}
\let\thanks\relax}

\def\@maketitle{
\hbox to\textwidth{\hfil
\hbox{\begin{tabular}{l}\@preprint\end{tabular}}}
\vskip 2em
\begin{center}{\Large\bf \@title \par}
\vskip 1.5cm {\normalsize
\lineskip .5em
\begin{tabular}[t]{c}\@author
\end{tabular}\par}
\end{center}\par\vskip 1.5em}

\def\preprint#1{\gdef\@preprint{#1}}

\def\abstract{\if@twocolumn
\section*{Abstract}
\else \normalsize
\begin{center}
{\large\bf Abstract\vspace{-.5em}\vspace{0pt}}
\end{center}\quotation\fi}
\def\endabstract{\if@twocolumn\else\endquotation\fi}
\catcode`\@=12

\def\be{\begin{equation}}
\def\ee{\end{equation}}
\def\bea{\begin{eqnarray}}
\def\eea{\end{eqnarray}}
\def\bdis{\begin{displaymath}}
\def\edis{\end{displaymath}}
\def\ba{\begin{array}}
\def\ea{\end{array}}
\def\bce{\begin{center}}
\def\ece{\end{center}}
\def\bfi{\begin{figure}}
\def\efi{\end{figure}}

\def\ggbmn{g_{\mu\nu}}
\def\atm{A^\mu}
\def\abm{A_\mu}
\def\atn{A^\nu}

\def\abr{A_\rho}
\def\bpsi{\bar{\psi}}

\def\ptm{\partial^\mu}
\def\pbm{\partial_\mu}

\def\pbn{\partial_\nu}

\def\cdbm{D_\mu}

\def\tint{\int d^2\! x\,\,}

\def\fsl#1{\setbox0=\hbox{$#1$}           
   \dimen0=\wd0                                 
   \setbox1=\hbox{/} \dimen1=\wd1               
   \ifdim\dimen0>\dimen1                        
      \rlap{\hbox to \dimen0{\hfil/\hfil}}      
      #1                                        
   \else                                        
      \rlap{\hbox to \dimen1{\hfil$#1$\hfil}}   
      /                                         
   \fi}
\newcommand{\dfrac}[2]{\frac{\strut \displaystyle{#1}}%
{\strut \displaystyle{#2}}}

\begin{document}
\baselineskip=.285in
\preprint{DPNU-94-52 \\ hep-th/9411201 \\ November 1994}
\title{ Thirring Model
 as a Gauge Theory\thanks{Submitted to Prog. Theor. Phys.} }
\author{
 Taichi ITOH,
 Yoonbai KIM,
 Masaki SUGIURA
 and Koichi YAMAWAKI\\ \vspace{6pt} \\
 {\it Department of Physics,
      Nagoya University,
      Nagoya 464-01 }\\  \\
 E-mail: taichi, yoonbai, sugiura,
 yamawaki$@$eken.phys.nagoya-u.ac.jp
       }
\maketitle
\renewcommand{\theequation}{\thesection.\arabic{equation}}

\abstract
%
We reformulate the Thirring model
in $D$ $(2 \le D < 4)$ dimensions as
a gauge theory by introducing $U(1)$ hidden local symmetry (HLS)
and study the dynamical mass generation of the fermion through
the Schwinger-Dyson (SD) equation.
By virtue of such a gauge symmetry we can greatly simplify
the analysis of the SD equation
by taking the most appropriate gauge
(``nonlocal gauge'') for the HLS.
In the case of even-number of (2-component) fermions,
 we find the dynamical
fermion mass generation as the second order
phase transition at certain fermion number, which breaks
the chiral symmetry but preserves the parity
 in (2+1) dimensions ($D=3$).
In the infinite four-fermion coupling (massless gauge boson) limit
in (2+1) dimensions,
the result coincides with that of the (2+1)-dimensional QED,
 with the critical
number of the 4-component fermion being
$N_{\rm cr} = \frac{128}{3\pi^{2}}$.
As to the case of odd-number (2-component) fermion
 in (2+1) dimensions,
the regularization
ambiguity on the induced Chern-Simons term may
 be resolved by specifying
the regularization so as to preserve the HLS.
Our method also applies to the (1+1) dimensions,
the result being consistent with the exact solution.
The bosonization mechanism in (1+1) dimensional
 Thirring model is
also reproduced in the context of dual-transformed
 theory for the HLS.
\newpage

\pagenumbering{arabic}
\thispagestyle{plain}

\setcounter{section}{1}
\setcounter{equation}{0}
\bce\section*{\large\bf $\S$ 1. Introduction}
\ece\indent\indent
%
 Fermion dynamical mass generation is the central
 issue of the scenario
of the dynamical electroweak symmetry breaking
such as the technicolor\cite{WS} and
the top quark condensate\cite{MTY}.
 Special attention has recently been paid to
the role of four-fermion interaction
in the context of walking technicolor\cite{Hol},
strong ETC technicolor\cite{MY}
and top quark condensate\cite{MTY}.
These models are based on the Nambu-Jona-Lasinio
 (NJL) model\cite{NJL}
of the scalar/pseudoscalar-type four-fermion interactions
combined with the gauge interactions, so-called gauged NJL model,
whose phase structure has been extensively studied through the
Schwinger-Dyson (SD) equation (see Ref.\cite{KTY}).
It has been shown\cite{KTY} that the phase structure of
 such a gauged NJL model
in (3+1) dimensions ($D=4$)
 is quite similar to that of the $D$ $(2<D<4)$ dimensional
four-fermion theory of scalar/pseudoscalar-type (without gauge
interactions)\cite{KY}, often
 called Gross-Neveu model,
which is renormalizable in $1/N$ expansion\cite{RWP}.

What about the four-fermion interaction of vector/axialvector-type?
Does it also give rise to the fermion mass generation?
Of course it can be transformed into
 the scalar/pseudoscalar-type interaction
through the Fierz transformation, but they are independent of
each other in the usual framework of $1/N$ expansion
 (see, e.g., Ref.\cite{BKY}).
 This type of
four-fermion interaction in fact has been studied
 in combination with the
scalar/pseudoscalar-type (``generalized NJL model''), which
is now a popular model as a low energy effective theory of QCD.
It is well-known that  at the usual $1/N$ leading
order in the ``generalized NJL model'' only the
scalar/pseudo\-scalar four-fermion interaction contributes to the
gap equation or to the fermion
mass generation, while  the vector/axialvector-type
 four-fermion interaction does not.
 Thus we may address the following question:
If the scalar/pseudoscalar-type four-fermion interaction does
not exist at all and the formal $1/N$ leading order is missing in
the above gap equation,
 is the fermion dynamical mass still generated by
the vector/axialvector-type alone?
If it is the case, such a dynamics
would be interesting for the model building beyond the
standard model.
It would also be interesting if there arises a situation
similar to the case of
scalar/pseudoscalar-type: Namely, the phase structure of
the gauged NJL model of vector/axialvector-type
might have some resemblance to
that of the  Thirring model\cite{Thi}
(without gauge interactions) in $D$ $(2<D<4)$ dimensions
which is known to be renormalizable in $1/N$ expansion\cite{RWP,KK}.

Thus we wish to study the fermion dynamical mass generation in
the $D$ $(2 \le D < 4)$ dimensional Thirring model
including the $D=2$ case.
The Thirring model
has been  extensively
studied  in (1+1) dimensions
since it is explicitly solvable\cite{Thi}.
However, it is only recent that the fermion dynamical mass generation
 in (2+1)-dimensional Thirring model has
been studied by several authors \cite{GMRS,HP,RS,HLL,ACCP}.
The results of these papers,
 however, are different from each other and rather
confusing partly due to lack of the discipline of the
analysis. In Refs.\cite{GMRS,HP},
for example, they introduced vector auxiliary field and pretended
 it as a gauge field despite the absence of manifest gauge symmetry.

In this paper we reformulate the Thirring
model as a gauge theory by introducing the
 hidden local symmetry (HLS)\cite{BKY} (see also \cite{GCS,KK}).
 When we fix the gauge of HLS to the unitary gauge, we get back to
 the original Thirring model written in terms of the vector auxiliary
 field. However, the unitary gauge is notorious for making
the actual analysis difficult, while
the existence of such a gauge symmetry has some virtues to make the
analysis consistent and systematic.

In the case of odd number of 2-component fermions, a peculiarity
arises in (2+1) dimensions, namely, the possibility of induced
Chern-Simons (CS) term\cite{RS}. We may take advantage of
existence of the gauge symmetry to resolve the problem of
regularization ambiguity concerning
 the induced CS term; the regularization must be chosen in
such a way as to keep the gauge symmetry (Pauli-Villars
 regularization)
as in the (2+1)-dimensional QED (QED${}_3$)\cite{Red}.
 The parity violating CS term
 will arise from the Pauli-Villars regulator even
 in the symmetric phase
where the fermion mass is not dynamically generated.

As to the case of even number of 2-component fermions,
 on the other hand,
we expect that parity violation in (2+1) dimensions
will not be induced, since the above induced CS term of each
fermion species can be arranged in pair of opposite sign
to cancel each other.
We shall demonstrate that an appropriate HLS
gauge fixing (nonlocal gauge\cite{GSC}) other
 than the unitary gauge
actually leads to the simple and consistent analysis
 of the SD equation
for the $D$ $(2 \le D < 4)$ dimensional Thirring model.
The nonlocal gauge is the gauge having
 no wave function renormalization
for the fermion and is the only way
 to make the bare vertex (ladder)
approximation to be consistent with
 the Ward-Takahashi (WT) identity
for the current conservation.
We find
dynamical chiral symmetry breaking which is parity-conserving in
(2+1) dimensions
in accord with the Vafa-Witten theorem\cite{VW} and
establish a second order phase transition
at a certain number of 4-component
fermions $N$ for each given value
 of the dimensionless four-fermion coupling $g$,
namely, the critical line on the $(N,g)$ plane.
This is somewhat analogous to the
existence of the critical $N$ in the QED${}_3$
\cite{MMV,N,KEIT}, which has been confirmed
 by the lattice Monte Carlo
simulation \cite{DKK}. Actually, when the four-fermion
coupling constant $g$ goes to infinity
 (massless gauge boson limit) for $D=3$,
our critical $N$ is explicitly evaluated as
$N_{\rm cr} =\frac{128}{3\pi^{2}}$
 in perfect agreement with that of the
QED${}_3$ in the nonlocal gauge\cite{N,KEIT}.
Note that this massless vector
limit is smooth thanks to the HLS in contrast to the original
Thirring model (unitary gauge) where this limit is singular and
ill-defined.
 For $D=2$ we explicitly solve the gap equation,
which turns out to be consistent with the exact solution.

Although at tree level the HLS gauge boson
 is merely the auxiliary
field, it is a rather common phenomenon
 that the HLS gauge boson acquires
kinetic term by the quantum corrections
 and hence becomes dynamical \cite{BKY}.
 We shall show that the HLS gauge boson also becomes
dynamical in the Thirring model once the fermion mass is
dynamically generated.
 In the infinite four-fermion coupling limit
this dynamical gauge boson becomes massless and the HLS becomes
a spontaneously unbroken gauge symmetry.
Were it not for the HLS, on the other hand,
this limit becomes ill-defined due to lack
of the manifest gauge symmetry \cite{BKY}.

Another advantage of the HLS is that
 we can use the dual transformation in
the Thirring model.
 Dual transformation is one which
manifests the propagating degrees of
 freedom and maps the theory with
strong gauge coupling to that
 with weak coupling constant \cite{Sav}.
In (1+1) dimensions the HLS together with the dual
transformation offers us
 a straightforward method for the bosonization
of the Thirring model in the context of path integral.

The rest of this paper is organized as follows.
In Section 2 we introduce
the model in $D$ $(2 \le D < 4)$ dimensions
and reformulate it by use of the hidden local
symmetry.
 The nonlocal gauge is introduced
 at the Lagrangian level rather than
in the Schwinger-Dyson equation,
 which makes the BRS invariance transparent.
In Section 3 we study the
SD equations under the nonlocal
 $R_{\xi}$ gauge fixing for HLS and
establish the existence of chiral symmetry breaking
dynamical mass generation and
of the associated
critical line on the $(N,g)$ plane
 in the case of even number of 2-component fermions.
In (2+1) dimensions this mass keeps the parity
 in accord with the Vafa-Witten theorem.
In (1+1) dimensions the fermion mass
 generation always takes place
as far as $g$ is positive.
 In Section 4 we demonstrate that
 the gauge boson of the HLS develops
a pole due to quantum correction (fermion loop)
 in the broken phase where
 the fermion mass is dynamically generated.
Section 5 is devoted to the dual transformation of the
Thirring model as a new feature of the HLS
and to the study of various
aspects of  the dual transformed theory,
 particularly the bosonization of
Thirring model in (1+1) dimensions.
We conclude in Section 6 with some discussions.
 In Appendix a proof is
given of the BRS invariance of the
 Thirring model Lagrangian with the
HLS in the nonlocal gauge.

\setcounter{section}{2}
\setcounter{equation}{0}
\bce\section*{\large\bf $\S$ 2. Hidden Local Symmetry}
\ece\indent\indent
%
In this section we introduce the
HLS \cite{BKY} into the massless
Thirring model in $D$ $(2 \le D<4)$ dimensions.
The Lagrange density of the Thirring model
is given by
\be
\label{thi}
{\cal L}_{\rm Thi}
= \sum_{a}\bar{\Psi}_{a}i\gamma^{\mu}\partial_\mu \Psi_{a}
-\frac{G}{2N}\sum_{a,b}
(\bar{\Psi}_{a}\gamma^{\mu}\Psi_{a})\,
(\bar{\Psi}_{b}\gamma_{\mu}\Psi_{b}),
\ee
where $\Psi_{a}$ is a 4-component
 Dirac spinor (although formal in $D$
dimensions) and $a,\,b$ are summed over
from 1 to $N$.
 Let us rewrite the theory by introducing an auxiliary
vector field $\tilde{A}_{\mu}$:
\be
\label{amu}
{\cal L^{\prime}}=
\sum_{a}\bar{\Psi}_{a}
i\gamma^{\mu}\tilde{D}_{\mu}\Psi_{a}+\frac{1}{2G}\tilde{A}_{\mu}
\tilde{A}^{\mu},
\ee
where $\displaystyle{ \tilde{D}_{\mu}
=\partial_\mu -\frac{i}{\sqrt{N}}
\tilde{A}_{\mu} }$.
Note that $\tilde{D}_{\mu}$ is not
 the covariant derivative in spite of its
formal similarity, since the field
 $\tilde{A}^{\mu}$ is just a vector field
 which  depicts
the fermionic current and does not transform as a gauge field.
Actually, the Lagrangian (\ref{amu}) has no gauge symmetry.
It is easy to see that when we solve away the auxiliary
field $\tilde{A}_{\mu}$
through the equation of motion for $\tilde{A}_{\mu}$,
Eq.(\ref{amu}) is reduced back
 to the original Thirring model (\ref{thi}).

Based on the ``$U(1)/U(1)$'' nonlinear sigma model,
we now show that Eq.(\ref{amu})
 is gauge equivalent to another model
possessing a symmetry
$U(1)_{\rm global}\times U(1)_{\rm local}$,
 with the $U(1)_{\rm local}$
being HLS \cite{BKY,GCS,KK}:
\bea
{\cal L_{\rm HLS}}&=&
\sum_{a}\bar{\psi}_{a}i\gamma^{\mu}D_{\mu}\psi_{a}
  -{N \over 2G}(D_{\mu} u\cdot u^{\dagger})^2 \nonumber \\
&=&\sum_{a}\bar{\psi}_{a}i\gamma^{\mu}D_{\mu}\psi_{a}
 +\frac{1}{2G}(A_{\mu}-\sqrt{N}\pbm\phi)^{2},
\label{hls}
\eea
where
 \be
 u=e^{i\phi}
 \ee
and
 $\displaystyle{ D_{\mu}=\partial_\mu
 -\frac{i}{\sqrt{N}}A_{\mu} }$
 is the covariant derivative for $A_{\mu}$
which is a gauge field in contrast to $\tilde{A}_{\mu}$ in
 Eq.(\ref{amu}).
 It is obvious that
Eq.(\ref{hls}) possesses a $U(1)$ gauge symmetry
 and is invariant
 under the transformation:
\be
\psi_a \mapsto \psi_a^{\prime} = e^{i\alpha}\psi_a, \;\;\;
A_\mu \mapsto A^{\prime}_{\mu}
=A_{\mu} + \sqrt{N}\partial_\mu \alpha, \;\;\;
\phi\mapsto \phi^{\prime}=\phi+\alpha.
\label{HLTransf}
\ee
Actually,  $\phi$ is the fictitious
 Nambu-Goldstone (NG) boson field
 which is to be absorbed into the longitudinal
 component of $A_{\mu}$.
If we fix the gauge by the gauge transformation
into the unitary gauge $\phi^{\prime}=0$ $(\alpha = -\phi)$:
\bea
\Psi_{a}&=&\psi_a^{\prime}=e^{-i\phi}\psi_{a},\\
\tilde{A}_\mu&=&A^{\prime}_\mu=A_\mu -\sqrt{N}\partial^{\mu}\phi,
\eea
then the Lagrangian (\ref{hls}) precisely
 coincides with Eq.(\ref{amu}).
Thus the original Thirring model is nothing but the gauge-fixed
(unitary gauge) form of our HLS model.
The mass of the vector boson $\tilde{A_{\mu}}$
 is now regarded as that of the
gauge boson $A_{\mu}$ generated
 through the Higgs mechanism \cite{BKY}.
This $U(1)$ case is actually identical with what is
known as the St\"uckelberg formalism
 for the massive vector boson.

There are several virtues of the existence of
 such a gauge symmetry:
First, the gauge symmetry enables us
 to prove straightforwardly the
S--matrix unitarity through the BRS symmetry
 (see Ref.\cite{KO}).
Secondly, actual calculations, particularly
 loop calculations,
 are generally hopeless in the unitary gauge,
 while the HLS provides us with
 the privilege to take the most appropriate
 gauge for our particular purpose.
Let us consider a general gauge $F[A]=0$ and
introduce the gauge fixing term into Eq.(\ref{hls}):
\be
{\cal L_{\psi,A}}=
\sum_{a}\bar{\psi}_{a}i\gamma^{\mu}D_{\mu}\psi_{a}
    + \frac{1}{2G}(A_{\mu}-\sqrt{N}\partial_{\mu}\phi)^2
    - \frac12 F[A] \left( \frac{1}{\xi(\partial^2)} F[A] \right) ,
\ee
where, by introducing the momentum- (derivative-)
 dependence of the
gauge fixing parameter $\xi$, we have formulated
 at Lagrangian level the
so-called nonlocal gauge\cite{GSC}
 which has been discussed only at the
SD equation level.
The covariant gauge is given by
 $F[A] = \partial_{\mu} A^{\mu}$ in which
the fictitious NG boson $\phi$ is not decoupled (except the
Landau gauge $\xi=0$).

More interesting gauge is the $R_{\xi}$ gauge\cite{FLS},
\be
F[A] = \partial_{\mu}A^{\mu}
 +\sqrt{N} \frac{\xi(\partial^2)}{G}\phi,
\label{Rxi}
\ee
which can again be a nonlocal gauge through
 the dependence of $\xi$ on
the derivative. The gauge fixing term
 in the nonlocal $R_{\xi}$ gauge
is given by
\be
\label{GF}
{\cal L}_{\rm GF}
 = -\frac12 \left(
             \partial_{\mu}A^{\mu}
 +\sqrt{N} \frac{\xi(\partial^2)}{G}\phi
            \right)
            {1 \over \xi(\partial^2)}
            \left(
             \partial_{\nu}A^{\nu}
 +\sqrt{N} \frac{\xi(\partial^2)}{G}\phi
            \right).
\ee
Putting Eq.(\ref{hls}) and Eq.(\ref{GF}) together, we arrive at
the Lagrangian in the nonlocal $R_{\xi}$ gauge:
\bea
 {\cal L} &=& {\cal L}_{\psi, A}+{\cal L}_{\phi},
  \label{total} \\
 {\cal L}_{\psi,A}
     &=& \sum_{a} {\bar \psi}_{a}i\gamma^{\mu} D_{\mu} \psi_{a}
               + \frac{1}{2G}(A_{\mu})^2
                 - \frac12 \partial_\mu A^\mu
                   \left(
                    \frac{1}{\xi(\partial^2)} \partial_\nu A^\nu
                   \right),
 \label{gf} \\
 {\cal L}_{\phi} &=& \frac{1}{2}
   \partial_{\mu} \phi \partial^{\mu} \phi
    - \frac{1}{2G}(\xi(\partial^2)\phi) \phi,
\eea
where we have rescaled the $\phi$ as
 $\sqrt{N/G}\phi \mapsto \phi$.
Thus the fictitious NG boson $\phi$
 is completely decoupled independently
 of $\xi(\partial^2)$ in the
 $R_{\xi}$ gauge, whether $\xi$ is nonlocal or not.
It is straightforward to prove that
 the above Lagrangian (even
in the nonlocal gauge) possesses
 the BRS symmetry (see Appendix).
Eq.(\ref{gf}) might appear as if we added
 the ``covariant gauge fixing term''
to  Eq.(\ref{amu}), although there is no gauge symmetry
and $\tilde{A}_{\mu}$ is not a gauge field in Eq.(\ref{amu}).
Such a confusion was actually made
 by some authors\cite{GMRS,HP}
who happened to arrive
at the Lagrangian having the same form as Eq.(\ref{amu}) in
the case of constant $\xi$.
This Lagrangian (\ref{gf}), whether the gauge parameter
 is nonlocal or not,
 can only be justified through the HLS in
the $R_{\xi}$ gauge. If we take the covariant gauge,
 on the other hand,
the field $\phi$ does not decouple
 except for the Landau gauge as
we have already mentioned.

In the next section we shall demonstrate that
the analysis of the fermion dynamical mass generation in
the ladder SD equation can be greatly simplified
by taking the nonlocal gauge\cite{GSC}
 of the $R_{\xi}$ gauge for HLS.
The nonlocal gauge is the gauge in which the fermion gets
no wave function renormalization, i.e., $A(-p^2)=1$
 for the fermion propagator $iS^{-1}(p)=A(-p^2)\fsl{p}-B(-p^2)$.
This gauge is
 necessary for the ladder (bare vertex) approximation
to be consistent with the WT identity for the $U(1)$ gauge
symmetry (or the current conservation of the global $U(1)$
 symmetry in the original Thirring model), which actually
requires no wave function renormalization.
In the usual gauge ($\xi=$ constant) including the Landau
gauge ($\xi=0$), on the other hand,
we cannot arrange $A(-p^2)=1$ without modifying the bare vertex
 into a complicated one consistent with the WT identity
in rather arbitrary way.

Thirdly, the massless vector boson limit
(limit of infinite four-fermion coupling constant) can be taken
smoothly in the gauges other than the unitary gauge
(original Thirring model) in the HLS formalism,
 so that our result in (2+1) dimensions
 can be compared with QED${}_3$, which is impossible in
the unitary gauge.
Such a massless limit is also interesting in the
composite models of gauge bosons\cite{BKY,GCS,SHK}.

Fourthly, the HLS can also be used to settle
 the regularization ambiguity
on the induced CS term in (2+1)-dimensional Thirring model.
Without gauge symmetry, any regularization
could  be equally allowed, which then
leads to contradictory result on whether
or not the CS term is induced by the fermion loop.
Once the HLS is explicit,
 regularization must be such as to preserve
the HLS (Pauli-Villars regularization),
which then concludes that the CS term is actually induced
in the same way as in QED${}_3$\cite{Red}.
Then there exists CS term  for the
 odd number of 2-component fermions even in the symmetric phase
 where the fermion mass is not generated,
 while for the even number of
 2-component fermions it can be arranged to
 cancel each other within the pairs of the regulators.

At this point one might still
suspect that the HLS is just a redundant degree of
freedom and plays no significant role on physics,
since there is no kinetic term for $A_{\mu}$  at tree level.
However, as is well known in the
$D$ $(2 \le D<4)$-dimensional
 nonlinear sigma model like $CP^{N-1}$ model,
the HLS gauge boson acquires kinetic term
through loop effects\cite{BKY}.
 Moreover, there exist realistic examples
of such dynamical gauge bosons of HLS realized in Nature:
The vector mesons ($\rho, \omega,...$)
 are successfully described as
the dynamical gauge bosons of the HLS in the nonlinear chiral
Lagrangian \cite{BKUYY,BKY}.
We shall demonstrate in the next section that
in the case at hand this phenomenon actually takes place,
once the fermion gets dynamical mass from the
nonperturbative loop effects in the SD equation.
In passing, the limit of infinite four-fermion coupling can
be taken only through the HLS, namely, the
 massless gauge boson can be generated
dynamically only when
the manifest gauge symmetry does exist \cite{BKY}.

Our HLS Lagrangian can easily be extended
 to the non-Abelian case by
use of the ``$U(n)/U(n)$'' nonlinear sigma model
 which is gauge equivalent
to $U(n)_{\rm global}\times U(n)_{\rm local}$
 model\cite{BKY,GCS}:
\be
\label{non}
{\cal L}=\sum_{a}\bar{\psi_a}i\gamma^{\mu}D_{\mu}
\psi_{a} -{N \over G}
 tr \left[ (D_{\mu} u\cdot u^{\dagger})^2 \right],
\ee
where
$A_{\mu}=A_{\mu}^{\alpha} T^{\alpha}$,
and $u=e^{i\phi}, \phi = \phi^{\alpha}T^{\alpha} $,
with $T^{\alpha}$ being the  $U(n)$ generators.
Actually, Eq.(\ref{non}) is gauge equivalent to
the Thirring model having the interaction
\be
-\frac{G}{2N}\sum_{a,b,\alpha}
(\bar{\Psi}_{a}\gamma^{\mu}T^\alpha \Psi_{a})\,
(\bar{\Psi}_{b}\gamma_{\mu}T^\alpha \Psi_{b}).
\ee
In contrast to the $U(1)$ case, however,
 the fictitious NG bosons
$\phi$ in the non-Abelian case are not
 decoupled even in the $R_{\xi}$ gauge,
which would make the SD equation analysis rather complicated.

\setcounter{section}{3}
\setcounter{equation}{0}
\bce\section*{\large\bf $\S$ 3. Schwinger-Dyson Equation}
\ece\indent\indent
%
In this section we study the fermion dynamical
 mass generation in the
$D$ $(2 \le D < 4)$-dimensional massless Thirring model through
the SD equation.
First of all, we must clarify what symmetry is to
be dynamically broken by the dynamical
 generation of fermion mass.
In this respect $D=3$ is rather special.
 Since we wish to study the fermion mass generation in
$D$ $(2 \le D < 4)$ dimensions which contain $D=3$,
 we here identify
the types of fermion mass and the symmetries
 to be broken in (2+1)
 dimensions \cite{ABKW}.

\setcounter{subsection}{1}
\bce\subsection*{\large\bf $\S$ 3.1. Chiral and parity symmetries
in $(2+1)$ dimensions}\par
\ece

In (2+1) dimensions the simplest representation of $\gamma$ matrices
 is the one with respect to the $2\times2$ matrices,
 or the Pauli matrices,

\be
\gamma^{0}=\sigma^{3},\;
\gamma^{1}=i\sigma^{1},\;
\gamma^{2}=i\sigma^{2}.\;
\ee
The 2-component fermions are denoted by $\chi_a$ with the
flavor index $a=1,...,N$. The parity transformation is defined by
\be
\chi_{a}(x)\mapsto\chi_{a}^{'}(x^{'})
=e^{i\delta}\sigma^{1}\chi_{a}(x)\;\;\;
\mbox{for} \;x^{'}=(t,-x,y).
\ee
Note that the mass terms for 2-component fermion,
$m\bar{\chi_a}\chi_a$, are odd under the parity symmetry.
We are restricting
 our interest to even number of fermion species
 and addressing  ourselves
to the question whether the symmetries of
 the classical Lagrangian are preserved
at quantum level or not.
In this case it is convenient to write the
 theory in terms of
4-component spinors
 $\psi_{a}\equiv\biggl(\ba{c} \chi_{a}\\ \chi_{N+a}
\ea\biggr)$ with flavor index $a=1,...,N$
 as we did in Section 2.
 The three $4\times 4\;\gamma$ matrices can be taken to be
\be
\gamma^{0}=\biggl(\ba{cc} \sigma_{3}&0\\ 0&-\sigma_{3}\ea\biggr),\;
\gamma^{1}=\biggl(\ba{cc} i\sigma_{1}&0\\ 0&-i\sigma_{1}\ea\biggr),\;
\gamma^{2}=\biggl(\ba{cc} i\sigma_{2}&0\\ 0&-i\sigma_{2}\ea\biggr),
\ee
and then there are three more $4\times 4$ matrices
\be
\gamma^{3}=\biggl(\ba{cc}0&1\\ -1&0\ea\biggr),\;
\gamma^{5}\equiv i\gamma^{0}\gamma^{1}\gamma^{2}\gamma^{3}=
\biggl(\ba{cc}0&1\\ 1&0\ea\biggr),\;
\tau\equiv -\gamma^{5}\gamma^{3}=\biggl(\ba{cc}1&0\\
 0&-1\ea\biggr),
\ee
which, together
with the identity, constitute generators of the global $U(2)$
symmetry, i.e.,
$\Sigma^{0}=I,\;\Sigma^{1}=-i\gamma^{3},\;
\Sigma^{2}=\gamma^{5},\;
\Sigma^{3}=\tau$.
Then the (2+1)-dimensional
massless Thirring model is invariant under the so-called
``chiral" transformation
\be
\psi_{a}\mapsto\psi^{'}_{a}=(U\psi)_{a},\; U\equiv\exp\biggl(
i\omega^{i\alpha}\frac{\Sigma^{i}}{2}\otimes T^{\alpha}\biggr),
\ee
where $T^{\alpha}$ denote the generators of $U(N)$, so
that the full chiral symmetry of
 the theory is $U(2N)$ as expected.
The parity transformation
 for 4-component fermions is composed of
that of 2-component fermions
and the exchange between the upper and
the lower 2-component fermions, specifically
\be
\psi_{a}(x)\mapsto\psi_{a}^{'}(x^{'})
=i\gamma^{3}\gamma^{1}\psi_{a}(x),
\ee
and the corresponding operation on the gauge field becomes
\be
\abm(x)\mapsto\abm^{'}(x^{'})=(-1)^{\delta_{\mu 1}}\abm(x).
\ee

Now we identify the peculiarity of $D=3$ dimensions;
the full global symmetries of the
Lagrangian (\ref{thi}) (or equivalently
Eq.(\ref{hls})) are the  parity and the global $U(2N)$ chiral
symmetry.
 Accordingly, in (2+1)-dimensional case,
 order parameter of the chiral
symmetry breaking  $U(2N) \rightarrow U(N)\times U(N)$
 is the parity-invariant mass defined by
\be
m\bar{\psi}_{a}\psi_{a}
=m\bar{\chi}_{a}\chi_{a}-m\bar{\chi}_{N+a}\chi_{N+a},
\ee
while order parameter of the parity symmetry
 breaking is given by another
type of mass term
\be
m\bar{\psi}_{a}\tau\psi_{a}=m\bar{\chi}_{a}\chi_{a}
+m\bar{\chi}_{N+a}\chi_{N+a}.
\ee

Though at this stage
we do not yet know whether the dynamical
 symmetry breaking really occurs
or not, we know what the breaking pattern
should be once it happens,
 thanks to
the Vafa-Witten theorem\cite{VW}. Namely, since
the tree-level gauge action corresponding to
 Eq.(\ref{hls}) is real and positive
semi-definite in Euclidean space, energetically favorable is
a parity conserving configuration consisting of
half the 2-component fermions acquiring equal positive masses
and the other half equal negative masses.
Such a parity-conserving mass is indeed
generated, as we shall show through the SD equation.
It  was also confirmed in QED${}_3$ where
the classical action shares the same structure as ours
except for the kinetic term of the gauge field,
 both  satisfying condition of the real positivity
 in Euclidean space \cite{ABKW}.
Moreover, the parity violating pieces including the
induced CS term\cite{Red}
do not appear in the gauge
sector whenever the number of 2-component fermions is even.
According to the above arguments, the pattern of
 symmetry breaking
we shall consider is not for the parity
 but for the chiral symmetry.
Thus we investigate the dynamical mass
 of the type $m\bar\psi \psi$
in the SD equation. In $D(\ne 3)$ dimensions,
 on the other hand,
 such a mass breaks the
chiral $U(N)\times U(N)$ symmetry of Eq.(\ref{amu})
(also Eq.(\ref{hls}))
down to
a diagonal $U(N)$ symmetry.
 Incidentally, the $U(1)$ subgroup
of this diagonal $U(N)$ was actually
 enlarged into the $U(1)_{\rm global}
\times U(1)_{\rm local}$ by the HLS in Section 2.
(See Ref.\cite{BKY}.)

\setcounter{subsection}{2}
\bce\subsection*{\large\bf $\S$ 3.2. Schwinger-Dyson
 equation in the nonlocal $R_{\xi}$ gauge}
\ece

Taking the above arguments into account,
 we now study the SD equation
 to confirm whether the chiral symmetry
 is spontaneously broken or not
 in the $D$ $(2\le D<4)$ dimensional Thirring model.
 We write the full fermion propagator
 as $S(p)=i[A(-p^2)\fsl{p}-B(-p^2)]^{-1}$,
 with $B$ being the order parameter
 of the chiral symmetry which preserves
 the  parity in (2+1) dimensions.
Then the SD equation for Eq.(\ref{gf})
 is written as follows:
\be
(A(-p^2)-1)\fsl{p}-B(-p^2)
=-\frac{1}{N}\int \frac{d^D q}{i(2 \pi)^D}
  \gamma_{\mu}\dfrac{A(-q^2)\fsl{q}
+B (-q^2)}{A^2(-q^2)q^{2}-B^{2}(-q^2)}
 \Gamma_{\nu}(p,q)\;iD^{\mu \nu}(p-q),
\label{eqn:sd}
\ee
where $\Gamma_{\nu}(p,q)$ and
 $D_{\mu \nu}(p-q)$ denote the full vertex function and
 the full gauge boson propagator,
 respectively.
We should apply some appropriate
 approximations to this equation
 so as to reduce it to the solvable integral equation
 for the mass function $M(-p^2) = B(-p^2)/A(-p^2)$.

 Following the spirit of the analysis
of QED${}_3$\cite{ABKW}, we here
adopt an approximation based on the large $N$ arguments,
 in which
$\Gamma_{\nu}(p,q)$ and $D_{\mu \nu}(p-q)$
 are those at the $1/N$ leading
order, namely, the bare vertex and
 the one-loop vacuum polarization
 of massless fermion loop, respectively:
\be
\Gamma_{\nu}(p,q) = \gamma_{\nu}  ,
\ee
\be
-iD^{\mu\nu}(k) = d(-k^2)\left(g^{\mu\nu}-\eta(-k^2)
                     \frac{k^\mu k^\nu}{k^2}\right),
 \label{eqn:apro1}
\ee
\be
 d(-k^2)=\dfrac{1}{G^{-1}-\Pi(-k^2)},\;\;\;
 \eta(-k^2)
=\dfrac{\xi(-k^2)\;\Pi(-k^2)-k^2}{\xi(-k^2)\;G^{-1}-k^2},
 \label{eqn:apro2}
\ee
where we have adopted a nonlocal $R_{\xi}$ gauge, Eq.(\ref{GF}),
 with
the momentum-dependent gauge parameter $\xi(-k^2)$, and
$\Pi(-k^2)$ is the one-loop vacuum polarization of massless
fermions:
\be
 \Pi^{\mu\nu}(k)
 = \left(g^{\mu\nu}-\frac{k^\mu k^\nu}{k^2} \right) \Pi(-k^2),
 \label{VPT}
\ee
which is readily calculated to be:
\be
 \Pi(-k^2)=-\frac{2\;tr\;I}{(4\pi)^{D/2}}
             \Gamma\left(2-\textstyle{\frac{D}{2}}\right)
                B\left(\textstyle{\frac{D}{2}},\;
                       \textstyle{\frac{D}{2}}
                 \right)(-k^2)^{D/2-1},
 \label{PI}
\ee
with $tr\;I$ being the trace of unit matrix in spinor indices
($4$ for $2<D<4$ and $2$ for $D=2$).

 Then the SD equation (\ref{eqn:sd}) is reduced
 to the following coupled equations for $A(-p^2)$ and $B(-p^2)$ :
\bea
A(-p^2)-1 & = & \dfrac{1}{N p^2}\int \frac{d^D q}{i(2 \pi)^D}
                \frac{A(-q^2)}{A^2(-q^2)q^2-B^2(-q^2)} \nonumber \\
          &   & d(-k^2)\left[\{ \eta(-k^2)+2-D\}(p\cdot q)
            -\dfrac{2 (k\cdot p)(k\cdot q)}{k^2}\eta(-k^2) \right],
\label{sda} \\
B(-p^2) & = & -\frac{1}{N} \int \frac{d^D q}{i(2 \pi)^D}
       \frac{B(-q^2)}{A^2(-q^2)q^2-B^2(-q^2)}d(-k^2)[D-\eta(-k^2)],
\label{sdb}
\eea
where $k_\mu =p_\mu -q_\mu$. These are our basic equations.

At first sight Eqs.(\ref{sda}), (\ref{sdb})
 might seem to be trivial in $1/N$ expansion,
since L.H.S. is formally of $O(1/N)$ whereas R.H.S. is of $O(1)$,
 which then would imply a trivial solution,
 $A(-p^2)=1$ and $B(-p^2)=0$, at the $1/N$ leading order.
 However, as was realized in  QED${}_3$\cite{ABKW},
 these equations are
 self-consistent nonlinear equations
 through which $A(-p^2)$ and $B(-p^2)$ may arrange
 themselves to balance the $N$-dependence of L.H.S.
 and that of R.H.S.
 in a nontrivial manner.
 In fact, $N$-dependence of the solution
 might be non-analytic in $N$ as was the case in
  QED${}_3$\cite{MMV}.
 We wish to find such a nonperturbative nontrivial solution
  by just examining the Eqs.(\ref{sda}), (\ref{sdb})
 for finite $N$.

A technical issue to solve Eqs.(\ref{sda}), (\ref{sdb}) is
how to handle the coupled SD equations Eqs.(\ref{sda}),
 (\ref{sdb}).
Here we follow the nonlocal gauge proposed
 by Georgi et al.\cite{GSC,KEIT}
which reduces the coupled SD equations
 into a single equation for $B(-p^2)$
by requiring $A(-p^2)\equiv 1$ in Eq.(\ref{sda})
 by use of the freedom
of gauge choice. This is actually the gauge
 in which the bare vertex approximation
can be consistent with the  WT identity
 for HLS (or the current conservation), i.e., $A(0)=1$.
This is in sharp contrast to the ordinary
(momentum-independent)
covariant gauge or  even the Landau gauge with
 $\eta(-k^2)=1$ ($\xi(-k^2)=0$),
in which the bare vertex approximation is not
consistent with the WT identity.
In the nonlocal gauge $B(-p^2)$ itself is a mass
 function, i.e.,
 $M(-p^2)= B(-p^2)$.

 Requiring $A(-p^2)\equiv 1$ in the nonlocal gauge,
 we perform the angular integration
of Eqs.(\ref{sda}), (\ref{sdb}) in Euclidean space
 (hereafter in this section we use the Euclidean notation):
\bea
0 & = &
        \int_0^{\pi} d\theta \sin^D \theta \left[
        \frac{1}{D-1}\dfrac{d}{d k^2}\left\{d(k^2)
(\eta(k^2)+D-2)\right\} +
        \dfrac{\eta(k^2) d(k^2)}{k^2}
        \right],\label{sdc} \\
B(p^2) & = & \dfrac{1}{N} \int_0^{\Lambda^{D-2}}
 d(q^{D-2}) K(p,q;G)
             \dfrac{q^2 B(q^2)}{q^2+B^2(q^2)},
             \label{sdd}
\eea
where we have introduced ultraviolet (UV)
 momentum cutoff $\Lambda$
 and the kernel $K(p,q;G)$ is given by
\be
K(p,q;G)={\textstyle \frac{1}{ (D-2) 2^{D-1} \pi^{(D+1)/2}
 \Gamma
 \left( \frac{D-1}{2} \right) } }
         \int_0^\pi d\theta \sin^{D-2} \theta \;
         d(k^2)[D-\eta(k^2)],
         \label{ker}
\ee
with $k^2 = p^2 + q^2 - 2 pq \cos \theta$.
It is
 easily seen that  Eq.(\ref{sdc}) is used to  determine
 the gauge fixing function $\eta(k^2)$:
\be
\eta(k^2)=(D-2)\left[\dfrac{D-1}{k^{2(D-1)}
           d(k^2)}\int_0^{k^2} d\zeta\;
\zeta^{D-2} d(\zeta) -1\right].
           \label{eta}
\ee
Substituting $d(k^2)$ in Eq.(\ref{eqn:apro2})
 into the above relation, we determine $\eta(k^2)$:
\bea
\eta(k^2)&=&(D-2)\left[\left(1+ \frac{Gk^{D-2}}{C_D} \right)
            {}_{2}F_{1}
            \left( \textstyle{ 1,1+\frac{D}{D-2},
2+\frac{D}{D-2};
             -\frac{Gk^{D-2}}{C_D} }
            \right)
            -1\right],\label{eqn:eta2} \\
 C_D^{-1} &\equiv& \frac{2\;tr\;I}{(4\pi)^{D/2}}
             \Gamma\left(2-\textstyle{\frac{D}{2}}\right)
                B\left(\textstyle{\frac{D}{2}},\;
                       \textstyle{\frac{D}{2}}
                 \right),
 \nonumber
\eea
with ${}_2 F_1(a,b,c;z)$ being the hypergeometric function.
Eq.(\ref{eqn:eta2}) actually forces the gauge fixing parameter $\xi$
 to be a function of $k^2$.
Once we determined $\eta(k^2)$ and hence the kernel
 Eq.(\ref{ker}),
 our task is now reduced to solving
a single SD equation for the mass function $B(p^2)$,
 Eq.(\ref{sdd}),
 which is much more tractable.

The kernel $K(p,q;G)$ in Eq.(\ref{ker})
 is a positive function for positive arguments $p$, $q$,
 and $G$, since $D-\eta(k^2)$ is positive for positive $k^2$.
Moreover, the kernel depends on the arguments $p$ and $q$
 only through $k^2 = p^2 +q^2 - 2pq \cos \theta$,
 so that it  has the symmetry under the exchange of $p$ and $q$.
 These kinematical properties of the kernel $K(p,q;G)$
 are essential to proving the existence of a nontrivial solution
 of the SD equation (\ref{sdd}) in the next subsection.

\setcounter{subsection}{3}
\bce\subsection*{\large\bf $\S$ 3.3. Existence of nontrivial solution
 and critical line}
\ece

 Now we investigate existence of the nontrivial solution for the
 SD equation (\ref{sdd}) for $2 \le D < 4$,
 based on the method of Refs.\cite{MN,Atk}.
In $D=2$ dimensions the SD equation
 reduces to the gap equation for the constant
 dynamical fermion mass as in the Gross-Neveu model.
We can explicitly solve this full gap equation.
 For $2 < D < 4$ we shall use the bifurcation method\cite{Atk}
 to solve the SD equation.

 Let us first consider $2 < D < 4$.
 The integral equation (\ref{sdd}) always has
 a trivial solution $B(p^2)\equiv 0$.
  We are interested in the vicinity
  of the phase transition point where the nontrivial solution
 also starts to exist.
 Such a bifurcation point is identified by
 the existence of an infinitesimal solution
 $\delta B(p^2)$ around the trivial solution
 $B(p^2)\equiv 0$ \cite{Atk}.
   Then we obtain the linearized equation for $\delta B(p^2)$:
\be
\delta B(p^2) = \frac{1}{N} \int_{m^{D-2}}^{\Lambda^{D-2}}
 d(q^{D-2})
                 K(p,q;G) \delta B(q^2),
\label{eqn:bifur}
\ee
where we introduced the IR cutoff $m$.
 It is enough for us to show the existence of a nontrivial
 solution of the bifurcation equation (\ref{eqn:bifur}) \cite{MN}.
 Particularly, we can obtain the exact phase
 transition point where the bifurcation takes place.
 Since we normalize the solution
 as $m=\delta B(m^2)$, $m$ is nothing but the dynamically
 generated
 fermion mass.

Rescaling $p = \Lambda x^{\frac{1}{D-2}}$ and $\delta B(p^2)
 = \Lambda \Sigma (x)$,
we rewrite Eq.(\ref{eqn:bifur}) as follows:
\be
\Sigma (x) = \frac{1}{N} \int_{\sigma_m}^{1}
 dy \tilde{K}(x,y;g) \Sigma (y),
\label{eqn:rbifur}
\ee
where we introduced
 $\sigma_m = (m/\Lambda)^{D-2}$ ($0 < \sigma_m \leq 1$),
the dimensionless four-fermion coupling constant
 $g = G/\Lambda^{2-D}$
 and
\be
\tilde{K}(x,y;g) \equiv K(x^{\frac{1}{D-2}},y^{\frac{1}{D-2}};g).
\ee
As we mentioned in the end of the last subsection,
the kernel $\tilde{K}(x,y;g)$ is positive and symmetric:
\be\label{eqn:prok}
\tilde{K}(x,y;g)=\tilde{K}(y,x;g)>0,\;\;\;
\mbox{for $x$, $y$ and $g \ge 0$.}
\ee
This is the most important property
for the existence proof of the nontrivial
solution \cite{MN}.

Let us consider the linear integral equation:
\be
\phi (x) = \frac{1}{\lambda}\int_{\sigma_m}^{1}
 dy \tilde{K}(x,y;g) \phi (y),
\ee
whose eigenvalues and eigenfunctions are denoted
 by $\lambda_{n} (g,\sigma_m)$
 ($ | \lambda_{n} | \geq | \lambda_{n+1}|;\; n = 1,2,\ldots . $)
 and  $\phi_{n} (x)$, respectively.
The kernel $\tilde{K}(x,y;g)$ is a symmetric one
 and hence satisfies the following property:
\be
\sum_{n=1}^{\infty} \lambda^2_{n} (g,\sigma_m) =
\int_{\sigma_m}^{1} \int_{\sigma_m}^{1}
 dx dy [\tilde{K}(x,y;g) ]^2 < \infty.
\label{eqn:eigen}
\ee
The R.H.S. of Eq.(\ref{eqn:eigen}) gives
 the upper bound for each
 eigenvalue $\lambda_{n} (g,\sigma_m)$.
Furthermore, using the positivity of the symmetric kernel
 (see Eq.(\ref{eqn:prok})),
 we can prove that the maximal eigenvalue
 $\lambda_{1} (g,\sigma_m)$ is always positive
 and the corresponding eigenfunction $\phi_{1}(x)$
 has a definite sign (nodeless solution).

In the bifurcation equation (\ref{eqn:rbifur})
 this implies the following:
{\em If $N$ is equal to the maximal
 eigenvalue of the kernel
$\lambda_1 (\alpha,\sigma_m)$}:
 $N=\lambda_1 (\alpha,\sigma_m)$,
{\em then there exists a nontrivial nodeless solution
 $\Sigma(x)=\phi_{1}(x)$ besides a trivial one.}
$N=\lambda_{1} (g,\sigma_m)$ determines
 a line on $(N,g)$ plane
 which is specified by $\sigma_m$.
Hence the above statement means
each line with the parameter $\sigma_m$
 corresponds to the
dynamically generated mass
$m=\Lambda(\sigma_m)^{\frac{1}{D-2}}$.
 Now we introduce $N_{\rm cr} (g)$ defined by
\be
N_{\rm cr} (g) = \lambda_1 (g,\sigma_m \rightarrow 0).
\ee
As $\sigma_m$ approaches zero, the corresponding line
also approaches the critical line on $(N,g)$
 plane; $N=N_{\rm cr}(g)$.
Moreover, since $\lambda_1(g,\sigma_m)$
 is the maximal eigenvalue of the kernel,
 there is no non-zero solution for $N$
 larger than $\lambda_1(g,\sigma_m)$.
 Through these consideration we can conclude that
{\em if the inequality $N < N_{\rm cr}(g)$ is satisfied,
 then the fermion mass is dynamically generated.}
 Existence of the critical line,
$N=N_{\rm cr}(g)$ or $g=g_{\rm cr}(N)$,
 in the two-parameter space is somewhat analogous
to that in the gauged NJL model\cite{KMY}.

Although it is difficult to obtain the explicit form
 of the critical line $N=N_{\rm cr}(g)$
 in the general case,
 we can do it
 in the limit of infinite four-fermi coupling constant,
 $g\rightarrow \infty$.
Let us discuss the (2+1) dimensions for definiteness,
 in which case the kernel reads
\be
 K(x,y;g \rightarrow \infty) = \frac{32}{3 \pi^2}\min \left\{
             \frac{1}{x},\;\;\frac{1}{y}\right\}.
\label{eqn:ker3}
\ee
Then the bifurcation equation (\ref{eqn:bifur})
 in (2+1) dimensions
 is rewritten into a differential equation
\be
\frac{d}{dx} \left(x^2 \frac{d\Sigma(x)}{dx}\right)
  =  -\frac{32}{3 \pi^2 N} \Sigma (x),
\label{Diffeqn}
\ee
plus boundary conditions
\bea
\Sigma^{\prime} (\sigma_m) & = & 0,
\;\;\mbox{(IR B.C.)}
 \label{eqn:irbc} \\
\mathop{[} x \Sigma^{\prime} (x)
+ \Sigma (x) \mathop{]} {}_{x=1}
 & = & 0.
\;\;\mbox{(UV B.C.)}
\label{eqn:uvbc}
\eea
Eqs.(\ref{Diffeqn}), (\ref{eqn:irbc}) and (\ref{eqn:uvbc})
 are the same as those in QED${}_3$ \cite{ABKW,N,KEIT}.
When $N > N_{\rm cr} \equiv 128/3 \pi^2$,
 there is no nontrivial solution of Eq.(\ref{Diffeqn})
 satisfying the boundary conditions,
 while for $N < N_{\rm cr}$ the following
bifurcation solutions exist:
\bea
\Sigma (x) &=& \frac{\sigma_m}{ \sin( \frac{\omega}{2} \delta) }
 \left( \frac{x}{\sigma_m} \right)^{- \frac12 }
\sin \left \{  \frac{\omega}{2} \left[ \ln \frac{x}{\sigma_m}
 + \delta \right] \right\},
\label{eqn:sol} \\
 \omega &\equiv& \sqrt{N_{\rm cr}/N - 1},\;\;\;
 \delta \equiv 2 \omega^{-1} \arctan \omega, \nonumber
\eea
where $\sigma_m$ is given by the UV
 boundary condition (\ref{eqn:uvbc}):
\be
\frac{\omega}{2} \left[ \ln \frac{1}{\sigma_m}
 + 2 \delta \right]
 = n \pi,\;\;\;
\mathop{n = 1,2,...}\;\;.
\label{eqn:const}
\ee
The solution with $n=1$ is
 the nodeless (ground state) solution
whose scaling behavior is read from Eq.(\ref{eqn:const}):
\be
\frac{m}{\Lambda}
 =e^{2 \delta} \exp
 \left[-\frac{2 \pi}{\sqrt{N_{\rm cr}/N - 1}}\right].
\label{eqn:scal}
\ee
 The critical number $N_{\rm cr} = 128/3\pi^2$
 is equivalent to the one in QED${}_3$
 with the nonlocal gauge fixing\cite{N,KEIT}.

Here we discuss the reason why our bifurcation equation
 at $g \rightarrow \infty$ in (2+1) dimensions
 is the same as the one in QED${}_3$.
In the nonlocal gauge the SD equation
 for QED${}_3$ reads\cite{KEIT}
\bea
 \delta B(p^2)
 &=&\frac{1}{N}\int_0^{\alpha} dq
    K_{{\rm QED}}(p,q;\alpha) \delta B(q^2), \\
K_{\rm QED}(p,q;\alpha)&=&\dfrac{\alpha}{4\pi^2}
\int_0^{\pi}d\theta
  \sin\theta\;d_{\rm QED}(k^2)[3-\eta(k^2)],
\label{eqn:Qsd}
\eea
where the scale $\alpha$ is defined by $\alpha=N e^2$
 with the gauge coupling $e$
and the nonlocal gauge function $\eta(k^2)$
 is given by Eq.(\ref{eta})
with $d(k^2)$ replaced by $\alpha d_{\rm QED}(k^2)$.
It is known that contribution to the kernel comes mainly
 from the momentum region $k < \alpha$ \cite{ABKW}.
Noting that $\Pi(k^2) \sim k$ from Eq.(\ref{PI}),
we can expand $\alpha d_{\rm QED}(k^2)$ in $k/\alpha$:
\be
\alpha\; d_{\rm QED}(k^2)
=\dfrac{\alpha}{k^2 - \alpha\Pi(k^2)}
          =\frac{-1}{\Pi(k^2)}
           \left\{ 1+{\cal O}
 \left( \frac{k}{\alpha} \right) \right\}.
 \label{QEDcase}
\ee
On the other hand, $d(k^2)$ in our case
 becomes identical with the first term of Eq.(\ref{QEDcase})
 in the limit of $g \rightarrow \infty$:
\be
 d(k^2)=\dfrac{1}{\Lambda\;g^{-1} - \Pi(k^2)}
        \rightarrow \frac{-1}{\Pi(k^2)}.
\ee
In spite of the big difference in the general form,
 $\alpha d_{\rm QED}(k^2)$ at $k/\alpha \ll 1$
 and $d(k^2)$ at $g^{-1} \ll 1$ are both dominated by
 the same vacuum polarization $\Pi(k^2)$,
 which yields the same $\eta(k^2)$ and hence the same kernel.
 This is the reason for the coincidence
of the value of $N_{\rm cr}$
 with that of the QED${}_3$.

However,
 we should note an essential difference
 of our case from the QED${}_3$.
Since
 the asymptotic behavior of the HLS gauge boson propagator
 is $ \sim 1/k$,
 the loop integration appearing in R.H.S. of the SD equation
 (\ref{eqn:sd}) is logarithmically divergent.
 This is due to lack of the kinetic term of the HLS gauge boson
 at tree level,
 in contrast to the photon in QED${}_3$
 whose asymptotic behavior is $\sim 1/k^2$.
 Hence, in order to keep the integral
 in Eq.(\ref{eqn:sd}) to be finite,
 we must introduce the cutoff $\Lambda$,
 whereas in QED${}_3$ the gauge coupling
 constant $\alpha$ provides
 an intrinsic mass scale which plays
 a role of the natural cutoff \cite{ABKW}.
 Therefore, different from QED${}_3$,
 $\Lambda$ should be removed
 by taking the limit $\Lambda \rightarrow \infty$
 in such a way as to keep the physical quantity
 like the fermion dynamical mass to be finite.
 This procedure corresponds to
the renormalization a \`la Miransky
 proposed in the strong coupling QED${}_4$\cite{Mir}.
 This renormalization defines the continuum theory
 at the UV fixed point located at
 the critical line $N=N_{\rm cr}(g)$ or $g= g_{\rm cr}(N)$
 in much the same way as the gauged NJL model\cite{KTY}.


Next we discuss the $D=2$ case,
 in which $\eta(k^2)=0$ from Eq.(\ref{eta})
 and thereby the kernel does not depend on $p$.
Then the SD equation (\ref{sdb}) is written as
\be
 B(p^2)
  =\frac{1}{N(1+\pi/G)}\int_0^\Lambda
 dq \dfrac{q B(q^2)}{q^2+B^2(q^2)}.
 \label{D=2SDeqn}
\ee
The R.H.S. of Eq.(\ref{D=2SDeqn})
 has no $p$ dependence, so that
 the fermion mass function is just
 a constant mass $B(p^2) \equiv m$.
This salient feature can only be realized
under the nonlocal gauge
 $\xi(k^2) \ne$const. we have chosen.
Therefore, as in the Gross-Neveu model,
 the above equation gives the gap equation:
\be
 m = \frac{1}{N(1+\pi/G)}\int_0^\Lambda dq \dfrac{q m}{q^2+m^2},
\ee
which has a nontrivial solution $m \ne 0$ for arbitrary $N$
 when $G>0$ or $G<-\pi$:
\be
 \label{eqn:cri}
 \ln \left(1+\frac{\Lambda^2}{m^2}\right)
  =2N\left(1+\frac{\pi}{G}\right).
\ee

Now the $(N,G)$ plane is divided into three regions with $N >0$;
 1.$\;G > 0$, 2.$\; -\pi < G < 0$, and 3.$\;G < -\pi$.
When we take the continuum limit $m/\Lambda \rightarrow 0$
 as in the case of $2<D<4$,
 we reach the critical line $G=0$ and $N>0$
 which may be interpreted as a trivial UV fixed line.
On the other hand,
 $G=-\pi$ is only reached by $\Lambda/m \rightarrow 0$
 (maybe a nontrivial ``IR fixed line'')
 and has nothing to do with the continuum limit.
The $\beta$-function in the broken phase ($G>0$, $G<-\pi$)
 is given by
\be
 \beta_N(G) \equiv
  \left.
    \Lambda {\partial G(\Lambda) \over \partial \Lambda}
  \right|_N
  =  - {G^2 \over \pi N}\left[1-e^{-2N(1+{\pi \over G})}
                         \right].
 \label{betafunc}
\ee
The region 2 allows only the trivial solution $m=0$.
In the next section we shall derive the bosonization
 in (1+1) dimensions
 through the dual transformation for HLS and show that
the theory with $-\pi<G<0$ lies not in the symmetric phase
 but has no ground state so that the trivial solution
 $m=0$ depicts an unstable extremum.
It means that the theory only has a broken phase
 with $G>0$ or $G<-\pi$.
Therefore the results based on the SD equation
 perfectly coincide with the exact results obtained by
 operator methods\cite{Thi} and those in Section 5.
This agreement is quite encouraging
 for the reliability of the SD equation
 in $D$ $(2<D<4)$ dimensions
 as well as $D=2$.

\setcounter{section}{4}
\setcounter{equation}{0}
\bce\section*{\large\bf $\S$ 4. Dynamical Gauge Boson}
\ece\indent\indent
%
In this section we discuss the dynamical
 pole generation of the HLS gauge
 boson which is merely the auxiliary field at tree level.
It is obvious that in the chiral symmetric phase
 where the fermions
 remain massless,
 there is no chance for the HLS gauge boson to develop
 a pole due to
 fermion loop effect,
 since a massive vector bound state,
 if it is formed,
 should decay
 into massless fermion pair immediately.
In fact the gauge boson propagator in Eq.(\ref{eqn:apro1})
 and Eq.(\ref{eqn:apro2})
 with the contribution from the vacuum polarization
 of massless fermions in Eq.(\ref{VPT})
 has no pole in the time-like momentum region.
However, once the fermion acquires the mass,
 the HLS gauge boson propagator can have a pole structure
 due to the massive fermion loop effect\cite{BKY}.
We here discuss the vacuum polarization tensor
 of massive fermion loop in $D$ $(2 \le D < 4)$ dimensions.

At this point one might suspect that the fermion mass effect
 on the vacuum polarization
 tensor may affect the analysis of
 the SD equation in Section 3
 where the vacuum polarization tensor (\ref{VPT})
 was calculated by
 the massless fermion loop.
However,
 the fermion mass effect on the SD equation
 through the vacuum polarization tensor enters
 the kernel only as a linear or higher terms in $\delta B(-p^2)$
 in the bifurcation form of Eq.(\ref{eqn:bifur}).
There exists a linear term of $\delta B(-p^2)$
 already in the integral of Eq.(\ref{eqn:bifur}),
 so that the mass effect on the vacuum polarization tensor
 yields only higher order in $\delta B(-p^2)$
 and can be neglected
 in the bifurcation equation.
Thus our analysis in Section 3 is totally
 unaffected by inclusion
 of the dynamically generated fermion mass
 in the vacuum polarization tensor.

Suppose that the fermion acquire the dynamical mass
 $m=\delta B(m^2)$.
Disregarding the momentum-dependence of the mass function
for simplicity,
we can calculate the one-loop vacuum polarization tensor
 for the HLS gauge boson as
\be
 \Pi^{\mu\nu}(k)
  = \left( g^{\mu\nu} -\dfrac{k^\mu k^\nu}{k^2} \right)
 \Pi(-k^2),
\ee
where
\newcommand{\tr}{\mbox{\rm tr}}
\bea
 \Pi(-k^2)
  &=& \frac{2\; \tr \; I}{(4\pi)^{D/2}}
       \Gamma \left( 2-\frac{D}{2} \right)
        \int_0^1 d x \dfrac{x(1-x)k^2}{[m^2-x(1-x)k^2]^{2-D/2}}
 \nonumber \\
  &=& {\tr \; I \; \Gamma \left( 2-\frac{D}{2} \right)
        \over
       3(4\pi)^{D/2} }
      { k^2 \over m^{4-D} }
      \;{}_2F_1(2,2-\textstyle{{D \over 2}},\textstyle{{5 \over 2}}
       ;\textstyle{{k^2 \over 4m^2}}).
 \label{eqn:VacPol}
\eea

Since the function $d(-k^2)$ in Eq.(\ref{eqn:apro1})
 is defined by $d(-k^2)=[G^{-1}-\Pi(-k^2)]^{-1}$,
 the pole mass $M_V$ of the dynamical gauge boson,
 if it exists,
 is defined by the following equation:
\be
 G^{-1} = \Pi(-M_V^2),
 \;\;\; 0 \leq M_V^2 < 4m^2.
 \label{eqn:PoleDef}
\ee
 First of all we observe that $M_V \rightarrow 0$
 as $G \rightarrow \infty$
 for arbitrary $D$.
This limit is well-defined only
 through the introduction of HLS
 which becomes a spontaneously unbroken gauge symmetry.
On the other hand,
the original Thirring model corresponding
 to the unitary gauge of HLS
 becomes ill-defined in this limit.

It is easily found that when $k^2$ approaches $4m^2$,
 $\Pi(-k^2)$ diverges in $2 \le D \le 3$.
Thus
 the solution of Eq.(\ref{eqn:PoleDef}) exists
  in $2 \le D \le 3$
 for any magnitude of coupling constant $G$
 (although it should be stronger than the critical value
 over which the dynamical fermion mass is generated).
Once the fermion acquires the mass $m$ dynamically,
 the HLS gauge boson always develops a pole at $M_V<2m$.
Specifically in (1+1) dimensions the mass function
 is really  momentum-independent, $B(-p^2)\equiv m$,
 and the above
calculation becomes exact.
Then the HLS gauge boson has a pole
in the broken phase with $G>0$ :
\be
 {1 \over \pi}\left[
    { 4m^2 \over M_V \sqrt{4m^2 - M_V^2}}
     \tan^{-1} \sqrt{{M_V^2 \over 4m^2-M_V^2}} - 1
              \right] = G^{-1}.
\ee
In (2+1) dimensions the HLS gauge boson pole is given by
\be
 {2 \over 3\pi}\left[ {4m^2 + M_V^2 \over 2M_V} \tanh^{-1}
 {M_V \over 2m}
           - m \right] = G^{-1}.
\ee
In the case of $3<D<4$, on the other hand,
  the R.H.S. of Eq.(\ref{eqn:PoleDef}) remains finite
 even when $k^2 \rightarrow 4m^2$,
 then there exists a lower bound of $G$
 under which the HLS gauge boson propagator has no pole.
This lower bound of $G$ is determined by
\be
 G^{-1}_V \equiv \Pi(-4m^2)
           = {4 {\it tr} I \; \Gamma(2-D/2)
 \over 3(4\pi)^{(D/2)}}\;
     {}_2F_1(2,2- \textstyle{{D \over 2}},
                \textstyle{{5 \over 2}};1) \times m^{D-2}.
 \label{eqn:PoleCondition}
\ee

The intriguing feature of our case is that
 both $G_V$ and $m$ are related to
 a single coupling constant $G$.
This is in contrast to the massive Thirring model
 where the fermion mass is given by hand,
 and also to the mixed model of Gross-Neveu
 and Thirring model
 where the fermion dynamical mass and
the HLS gauge boson mass
 are separately determined
 by the Gross-Neveu coupling and
the Thirring coupling, respectively.

\setcounter{section}{5}
\setcounter{equation}{0}
\bce\section*{\large\bf $\S$ 5. Dual Transformation and Bosonization}
\ece\indent\indent
%
Now that we have reformulated
the Thirring model as a gauge theory,
we can further gain an insight
 into the theory by using a technique
inherent to the gauge theory, namely,
the dual transformation \cite{Sav}.
Let us rewrite the theory with HLS in Eq.(\ref{hls}) by
use of the dual transformation, which in (1+1) dimensions
leads to the bosonization
of Thirring model in the context of path integral.

We first consider the path integral
 for the Lagrangian (\ref{hls}),
\be
Z_{\rm HLS} = \int[d\abm][d\phi][d\bpsi_{a}][d\psi_{a}]
 \exp i\int d^{D}\!x\biggl\{
              \sum_{a}\bar{\psi}_{a}i\gamma^{\mu}D_{\mu}\psi_{a}
                 +\frac{1}{2G}(A_{\mu}-\sqrt{N}\pbm\phi)^{2}
                    \biggr\} \label{Zhls}.
\ee
Linearizing
the ``mass term" of gauge field by introducing an auxiliary field
$C_{\mu}$, we obtain a delta functional for $\pbm C^{\mu}$ through
an integration over the scalar field $\phi$ as follows:
\footnote{The scalar phase $\phi$ can
in fact be divided into two parts:
 $\phi=\Theta+\eta$,
 where $\Theta$ expressed by multi-valued function
 describes the topologically nontrivial sector, e.g.,
 the creation and annihilation of topological solitons,
 and $\eta$ given by single-valued function
 depicts the fluctuation around a given topological sector.
Inclusion of the topological sector $\Theta$ induces a topological
 interaction term \cite{Sav}, though we neglect $\Theta$
 contribution in this section, since we are interested in
 $\phi$ as the NG mode.}
\bea
\lefteqn{\int [d\phi]
          \exp i\int d^{D}\!x\frac{1}{2G}(\abm-\sqrt{N}\pbm\phi)^{2}
        }
 \nonumber\\
 &=& \int[d\phi][dC_{\mu}]
          \exp i\int d^{D}\!x\biggl\{
                              -\frac{1}{2}C_{\mu}C^{\mu}
                              +\frac{1}{\sqrt{G}}C_{\mu}
                               (\atm-\sqrt{N}\ptm\phi)
                             \biggr\} \\
 &=& \int[dC_{\mu}]\delta(\pbm C^{\mu})
          \exp i\int d^{D}\!x\biggl\{
                              -\frac{1}{2}C_{\mu}C^{\mu}
                              +\frac{1}{\sqrt{G}}C_{\mu}\atm
                             \biggr\},
\eea
where the auxiliary field $C_{\mu}$ at tree
level is nothing but the conserved current.
 If we pick up $C_{\mu}$ by use of
dual antisymmetric-tensor field
$H_{\mu_{1}\cdots\mu_{D-2}}$ of rank $D-2$,
 which satisfies the Bianchi identity,
we have the following relation
\be
 \int[dC_{\mu}]\delta(\pbm C^{\mu})\cdots
  =\int[dC_{\mu}][dH_{\mu_{1}\cdots\mu_{D-2}}]
    \delta(C_{\mu_{1}} - \epsilon_{\mu_{1}\cdots\mu_{D}}
           \partial^{\mu_{2}}H^{\mu_{3}\cdots\mu_{D}}) \cdots .
\ee
Substituting the above relation into the path integral
 and integrating out the auxiliary field $C_{\mu}$,
 then we have
\bea
 Z_{\rm Dual} &=& \int[dH_{\mu_{1}\cdots\mu_{D-2}}]
                  [d\abm][d\bpsi_{a}][d\psi_{a}]
                \exp i\int d^{D}\!x\biggl\{
                      \sum_{a}\bpsi_{a}i\gamma^{\mu}\cdbm\psi_{a}
 \nonumber\\
          & &\hspace{1.2cm}
              +\frac{(-1)^{D}}{2(D-1)}
                     H_{\mu_{1}\cdots\mu_{D-1}} H^{\mu_{1}\cdots\mu_{D-1}}
              +\frac{1}{\sqrt{G}}\epsilon^{\mu_{1}\cdots\mu_{D}}A_{\mu_{1}}
               \partial_{\mu_{2}}H_{\mu_{3}\cdots\mu_{D}}
                                   \biggr\}
 \label{deltaf} \\
          &=& \int[dH_{\mu_{1}\cdots\mu_{D-2}}][d\bpsi_{a}][d\psi_{a}]\;
               \delta(\sum_{a} \frac{1}{\sqrt{N}}
                      \bpsi_{a} \gamma^{\mu_{1}} \psi_{a}
                     +\frac{1}{\sqrt{G}}\epsilon^{\mu_{1}\cdots\mu_{D}}
                      H_{\mu_{2}\cdots\mu_{D}})
 \nonumber\\
          & & \hspace{1.3cm}
               \exp i\int d^{D}\!x\biggl\{
                     \sum_{a} \bpsi_{a} i \gamma^{\mu} \pbm \psi_{a}
                    +\frac{(-1)^{D}}{2(D-1)}
                     H_{\mu_{1}\cdots\mu_{D-1}}H^{\mu_{1}\cdots\mu_{D-1}}
                                  \biggr\},
 \label{eqn:dual}
\eea
where $H_{\mu_{1}\cdots\mu_{D-1}}
=\partial_{\mu_{1}}H_{\mu_{2}\cdots\mu_{D-1}}
-\partial_{\mu_{2}}H_{\mu_{1}\mu_{3}\cdots\mu_{D-1}}+\cdots+(-1)^{D}
\partial_{\mu_{D-1}}H_{\mu_{1}\cdots\mu_{D-2}}$.
The Lagrangian (\ref{eqn:dual}) describes $N$ ``free'' fermions and
a ``free'' antisymmetric tensor field of rank $D-2$
 which are, however,
 constrained through the delta functional.
 This implies that the dual field is actually
 a composite of the fermions.
In (2+1) dimensions the dual gauge field $H_{\mu}$ has the vector
structure as $\abm$ does, while in (3+1) dimensions it is
the second-rank anti-symmetric tensor field
 $H_{\mu\nu}$,
 {\rm i.e.},
 $H_{\mu\nu\rho} \sim \epsilon_{\mu\nu\rho\sigma}
 {\bar \psi} \gamma^\sigma \psi$.

Although the delta functional in Eq.(\ref{eqn:dual})
 tells us that the dual field is a composite of the fermions,
 it is difficult to read directly phase structure
 of the Thirring model in this formalism.
If we look at the tree level Lagrangian (\ref{eqn:dual})
 in (2+1) dimensions,
 it might seem that the dual gauge field $H_\mu$ is massless
 independently of the phase structure.
In order to understand the pole structure of
 the dual gauge field $H_\mu$,
 however, we have to take into account the quantum effect.
 For that purpose we may ignore the contributions from the fermion
  one-loop diagrams except for the vacuum polarization,
 since they generate only the self-interaction terms
 of the gauge field $A_\mu$
 or equivalently those of the dual gauge field $H_\mu$.
Therefore we compute the vacuum polarization in Eq.(\ref{deltaf}):
\bea
 Z_{\rm Dual} \approx \int [dH_\mu][dA_\mu]
   \exp i \int d^3x \bigg\{
         \!\!\! &-& \! \! \frac14 H_{\mu\nu}H^{\mu\nu}
         + {1 \over \sqrt{G}} \epsilon^{\mu\nu\rho}A_\mu
 \partial_\nu H_\rho
 \nonumber \\
          & &- \frac12 A_\mu \left(
            g^{\mu\nu}- { \partial^\mu \partial^\nu \over \partial^2}
                         \right) \Pi(\partial^2) A_\nu
                    \bigg\}.
\eea
Integrating out the HLS gauge field $A_\mu$,
 we obtain an effective Lagrangian for the dual gauge field $H_\mu$
 without interaction terms:
\be
 {\cal L}_{\rm H}
  = {1 \over 2G} H_\mu \left(
                         g^{\mu\nu}\partial^2
- \partial^\mu \partial^\nu
                       \right)
    [G-  \Pi^{-1}(\partial^2)] H_\nu.
\ee
If we compare the pole structure of $H_\mu$ with that of $A_\mu$
 in Eq.(\ref{eqn:PoleDef}),
 we easily find that the dual gauge field $H_\mu$ shares
 exactly the same pole structure with the gauge field $A_\mu$
 irrespectively of the phase.

In (1+1) dimensions the relation in the delta functional
in Eq.(\ref{eqn:dual})
 implies nothing but the bosonization of Thirring model
 in the scheme of path integral, i.e.,
 $
 \frac{1}{\sqrt{G}}\epsilon^{\mu\nu} \pbn H
 \approx \frac{-1}{\sqrt{N}} \bpsi_a \gamma^\mu \psi_a
 $.
Integrating the fermions in Eq.(\ref{deltaf}),
 we obtain an effective theory which consists of a pseudoscalar
 and a vector gauge fields:
\be
 Z_{2D}=\int[dH][d\abm] \left( \det i\fsl{D} \right)^N \hspace{1mm}
         \exp i\tint\biggl\{
                           \frac12 \ptm H\pbm H
                          +\frac{1}{2 \sqrt{G}}
 H \epsilon_{\mu\nu} F^{\mu\nu}
                    \biggr\}.\label{eqn:2D}
\ee
The second term of the action in
Eq.(\ref{eqn:2D}) is the (1+1) dimensional
 analogue of axion term which
is the interaction term between the scalar
 and the gauge fields and takes the form
$H\epsilon_{\mu\nu\rho\sigma}F^{\mu\nu}F^{\rho\sigma}$ in (3+1)
dimensions.
Though the computation of fermionic determinant with
regularization generates the Abelian chiral anomaly,
this problem is resolved by the constant
shift of scalar field $H$ in axion term.
Since the fermionic determinant
is computed in an exact form, i.e.,
 $$
 -i N \ln
    \frac{\det i\fsl{D}}{\det i\fsl{\partial}}
  = \frac{1}{2\pi}\tint\atm
 (\ggbmn-\frac{\pbm\pbn}{\partial^{2}}) \atn,
 $$
 the integration over $\abm$ gives a free massless scalar theory
 as the bosonized Thirring model
\be\label{boson}
 Z_{\rm boson}=\int[dH]\exp
i\tint\frac{1}{2}(1+\frac{\pi}{G})\pbm H\ptm H .
\ee
If $G$ is in the region $-\pi<G<0$,
 the energy per unit volume is unbounded below and hence
 the (1+1)-dimensional Thirring model with
 coupling constant $G$ ($G>0$ or $G<-\pi$)
 has only the broken phase via the fermion dynamical mass
 generation as in the (1+1)-dimensional Gross-Neveu model.

\setcounter{section}{6}
\setcounter{equation}{0}
\bce\section*{\large\bf $\S$ 6. Conclusion and Discussions}
\ece\indent\indent
%
In this paper we have studied the Thirring model
 in $D$ $(2 \le D<4)$ dimensions
 and proposed how to
 understand it as a gauge theory through the introduction
 of hidden local symmetry. The advantage of manifest gauge symmetry
 was to let the various nonperturbative
 approaches tractable and provide
 the consistent method to treat such problems.

In the case of $2N$ 2-component fermions (or equivalently $N$
 4-component Dirac fermions) we studied the dynamical symmetry
 breaking in the context of $1/N$ expansion. Since we had the
 manifest $U(1)$ gauge symmetry,
 we took a privilege to choose a nonlocal
 $R_{\xi}$ gauge, which greatly simplified the analysis
 of the SD equation.

By using the bifurcation technique,
 we found a second order phase transition
 at a certain number of $N$ and $g$,
 thus having established the existence
 of the critical line on the $(N,g)$ plane.
We also proved existence of the nontrivial solution rigorously.
The HLS gauge boson became massless
 in the $g \rightarrow \infty$ limit,
 where the SD equation was solved analytically in (2+1) dimensions,
 yielding $N_{\rm cr}= \frac{128}{3\pi^{2}}$
 in perfect agreement with $N_{\rm cr}$ in QED${}_3$.
This limit makes sense thanks to the HLS, in sharp contrast
 to the original Thirring model where this limit is ill-defined.

In (1+1) dimensions, on the other hand,
 fermion mass is always generated,
 no matter what value $N(>0)$ and $G=g(>0)$ might take:
The theory is in one phase (broken phase) for $G>0$
 as in the (1+1)-dimensional Gross-Neveu model.
Our result is consistent with the exact solution
 of the (1+1)-dimensional Thirring model.
In this case there is no regularization ambiguity,
 since the regularization must respect the HLS
 as in the massless Schwinger model.

The dynamical symmetry breaking
 in (2+1)-dimensional Thirring model with
 many flavors have previously been discussed
 in $1/N$ expansion\cite{GMRS,HP,RS,HLL,ACCP}.
Here we compare our results
 with those of the previous authors.
When the auxiliary vector field $\tilde{A}_\mu$ has been used,
 the authors in Ref.\cite{GMRS,HP} pretended it as
 a gauge field and added the ``gauge fixing'' term,
 though the model carries no gauge symmetry.
The hidden local symmetry we found in such model explains
 that the gauge fixing they chose
 is neither the ``over-gauge fixing''
 nor the Landau gauge but the $R_{\xi}$ gauge,
 so that the awkward procedure of
 ``the gauge fixing without gauge symmetry''
 is justified by the discovery of hidden local $U(1)$ symmetry.
The ladder approximation for the vertex
 leads to $A(-p^2)=1$ to be consistent with the WT identity,
 while it is not allowed under the Landau gauge.
It can only be realized through the nonlocal
 $R_{\xi}$ gauge we chose.

The dynamical gauge boson is generated
 when fermions get mass.
The result in $2 \le D \le 3$  has a novel feature.
The gauge boson pole is
 always developed independently of the coupling $G$,
 once the fermion acquires mass at $G>g_{\rm cr}/\Lambda^{D-2}$.
 For $3<D<4$, on the other hand,
 the gauge boson pole can be generated
 only for $G > G_V$ which may or may
 not be satisfied by the coupling larger than
 the critical coupling $g_{\rm cr}/\Lambda^{D-2}$.
It would be interesting to see the precise relation between the critical
 coupling for the dynamical gauge boson generation
 and that of the fermion dynamical mass generation in this case.

We rewrote Thirring model with hidden local symmetry
 in terms of dual field.
In (1+1) dimensions, we demonstrated that
this dual transformation based on the introduction
of hidden local symmetry is a straightforward way to arrive
at the bosonization of Thirring model.
In (2+1) dimensions, it was also shown that
both the HLS gauge field $A_\mu$ and dual gauge field $H_\mu$
share the same mass spectrum;
they are massless in symmetric phase and they have equal mass
in broken phase. This formulation might be useful also
in $D$ $(2<D<4)$ dimensions.

In (2+1) dimensions we assumed two well-known results
 such that, if the number of 2-component fermions is even
 and the classical Lagrangian is parity-even,
 the parity-violating sector is not induced
 both in the effective action for the gauge field \cite{Red}
 and in the pattern of dynamically generated
 fermion masses \cite{VW}.
Both of them are consequences of exact calculations and,
 of course, it is consistent with our results.
However, the previous
 papers\cite{GMRS,RS,HLL,ACCP} claimed
 appearance of parity-violating piece through quantum effect,
 which is opposed to the results of both Refs.\cite{Red,VW}
 and ours given in Section 3.

The classical action in Eq.(\ref{hls})
 for odd number of massless fermions is invariant
 under the $U(1)$ HLS gauge transformation
 and the parity in (2+1) dimensions,
 while in the quantized theory both
 of them cannot be preserved simultaneously.
Since the regularization is to be specified
 so as to keep the $U(1)$ HLS (Pauli-Villars regularization),
 there is no regularization ambiguity in our case
 and hence the parity-violating anomaly
 arises as in QED${}_3$ \cite{Red}.
Therefore, the gauge theory described
 by the effective gauge field action lies
 in Chern-Simons Higgs phase:
\be
\label{csh}
{\cal L}_{\rm eff}(\abm)= \frac{1}{4\pi}
\lim_{ M_{Reg} \rightarrow \infty}
                       \frac{M_{Reg}}{ \left| M_{Reg} \right| }
                       \epsilon^{\mu\nu\rho}\abm\pbn\abr
                        +\frac{1}{2G}(\abm-\pbm \phi)^{2},
\ee
 where $M_{Reg}$ is the mass of the Pauli-Villars regulator.
Now an emphasis is in order on an important role of
the fictitious NG boson field $\phi$ which is
 composed of the topologically nontrivial sector $\Theta$ and
 the smooth NG degree $\eta$ for a given topological sector
 $\Theta$ as mentioned in the previous section.
If we neglect the topological sector and consider only
 the smooth NG boson mode $\eta$,
 then the nonderivative gauge mass term
 dominates in the long range physics
 and the theory remains just in the topologically trivial sector
 of Chern-Simons Higgs phase
 which is governed by a parity-violating helicity one photon
 with mass $2\pi /G$\cite{DY}.
Inclusion of the topological sector under the guiding principle
 of the HLS gives rise to the generation of CS vortices and
 realizes an anyonic phase\cite{HKP}.
We may recall that the addition of the CS term
 to the (2+1)-dimensional QED changed the
 structure of phase transition to the first-order one \cite{KM}
 and this theory has also been a model
 describing anyonic superconductivity.
Then the subject of dynamical symmetry breaking
 in massless Thirring model for
 an odd number of fermions may generate intriguing results.

\bce\section*{\large\bf Acknowledgments}\ece
\indent\indent
We would like to thank K.-I. Kondo for very
 helpful discussions on the
 nonlocal gauge and SD equation. Thanks are also due to
 T. Fujita, Jooyoo Hong,  D. Karabali, Choonkyu Lee,
 P. Maris, V.P. Nair, Q. Park, A. Shibata
 and S. Tanimura for valuable discussions.
Y.K. is a JSPS Postdoctoral Fellow (No.93033)
 and also thanks Center for Theoretical Physics at
 Seoul National University for its hospitality.
 This work was supported in part by a Grant-in-Aid for Scientific
 Research from the Japanese
 Ministry of Education, Science and Culture (No. 05640339).

\bce\section*{\large\bf Note added}\ece
\indent\indent
After completion of our work
 we have received a recent paper by S. Hands,
 ``$O(1/N_f)$ corrections to the Thirring Model in $2<d<4$'', Wales
 Preprint, SWAT/94/47 (Nov., 1994), which addresses the problem of
 the renormalizability but not the dynamical mass generation of
 the fermion.
This paper has the same difficulties as those in the previous
 papers which are resolved by the introduction of
 the hidden local symmetry as we mentioned in the text.
\newpage

\appendix
\setcounter{section}{1}
\setcounter{equation}{0}
\bce\section*{\large\bf Appendix.  BRS Invariance in the Nonlocal Gauge}
\ece\indent\indent
%
\newcommand{\BRS}{\mbox{\boldmath{$\delta_B$}}}
In this Appendix we prove that the gauge fixing term Eq.(\ref{gf})
is BRS invariant even for the nonlocal gauge. It might be nontrivial
to see that the nonlocal gauge fixing at Lagrangian level actually works.

According to the HLS transformations Eq.(\ref{HLTransf}),
the  BRS transformation for each field is given by
\begin{eqnarray}
  \BRS \psi_a(x)   &=& {i \over \sqrt{N}} c(x) \psi_a(x),\nonumber \\
  \BRS A_\mu(x)    &=& \partial_\mu c(x), \nonumber \\
  \BRS \phi(x)     &=& {1 \over \sqrt{N}} c(x),
 \label{BRSTransf1}  \\
  \BRS c(x)        &=& 0, \nonumber \\
  \BRS {\bar c}(x) &=& iB(x), \nonumber \\
  \BRS B(x)        &=& 0, \nonumber
\end{eqnarray}
where $c(x)$ and ${\bar c}(x)$ are ghost fields,
 and $B(x)$ is the so-called Nakanishi-Lautrap field.
 Moreover, it is well known that the operator $\BRS$ is nilpotent:
\be
 \BRS^2 \ast = 0.
 \label{BRSTransf2}
\ee

 Following the  text book procedure, we have
  the gauge fixing term plus Fadeev-Popov ghost term:
\be
 {\cal L}_{\rm{GF}+\rm{FP}}
  = -i \BRS \left( {\bar c} f[A,\phi,c,{\bar c},B]\right).
 \label{GFTerm}
\ee
 Without knowing explicit  form of $f[A,\cdots]$,
 we easily see
 from the nilpotency,
 Eq.(\ref{BRSTransf2}),
 that ${\cal L}_{{\rm GF}+{\rm FP}}$
  is BRS-invariant.

Now we show that the nonlocal $R_{\xi}$
gauge fixing term (\ref{GF}) is obtained,
if we take
 $f[A,\cdots]$ as
\bea
 f[A,\cdots]
  &=& \partial_{\mu}A^{\mu} +\sqrt{N} \frac{\xi(\partial^2)}{G}\phi
      + \frac12 \xi(\partial^2) B \nonumber \\
  &=& F[A] + \frac12 \xi(\partial^2) B,
 \label{FixFunc}
\eea
with $F[A]$ being the same as Eq.(\ref{Rxi}).
In fact, substituting Eq.(\ref{FixFunc}) into Eq.(\ref{GFTerm}),  we
find that ${\cal L}_{{\rm GF}+{\rm FP}}$ takes the form:
\bea
 {\cal L}_{\rm GF}
   &=& B \left\{ F[A] + \frac12 \xi(\partial^2) B \right\},
  \label{GFGF} \\
 {\cal L}_{\rm FP}
   &=& i{\bar c} \left( \Box + {\xi(\partial^2) \over G} \right) c.
  \label{GFFP}
\eea
As a usual result for the Abelian gauge symmetry,
 FP ghost decoulpes from the system completely.
We then translate $B(x) = B^\prime(x) - \xi^{-1}(\partial^2) F[A]$
in Eq.(\ref{GFGF}) and integrate out $B^\prime(x)$, arriving finally at
\be
 {\cal L}^\prime_{\rm GF}
  = -\frac12 F[A] {1 \over \xi(\partial^2)} F[A],
 \label{GFFinal}
\ee
where we have used the following identity:
\be
 \int d^Dx f(x) \xi(\partial^2) g(x)
 = \int d^Dx \left( \xi(\partial^2) f(x) \right) g(x),
\ee
 which can easily be proved in momentum space.

Eq.(\ref{GFFinal}) is nothing but the nonlocal
 $R_{\xi}$ gauge fixing term in
Eq.({\ref{GF}).

\end{document}